\documentclass[12pt]{article}
\usepackage{amsmath,amssymb,graphicx,mathrsfs,hyperref,slashed,setspace}

\usepackage{float}
\usepackage{latexsym}
\usepackage{bm}
\usepackage{blindtext}
\hypersetup{
colorlinks=true,
linkcolor=blue,
filecolor=black,
urlcolor=black,
citecolor=blue,
}

\newcommand{\hide}[1]{}

\renewcommand{\baselinestretch}{1.5}

\newcommand{\be}{\begin{equation}}
\newcommand{\ee}{\end{equation}}
\newcommand{\bea}{\begin{eqnarray}}
\newcommand{\eea}{\end{eqnarray}}

\def\({\left(} \def\){\right)}

\begin{document}
\title{\vspace{-1.8in}
{Defrosting frozen stars: \\ spectrum of  internal fluid modes}}
\author{\large Ram Brustein${}^{(1)}$,  A.J.M. Medved${}^{(2,3)}$, Tom Shindelman${}^{(1)}$
\\
\vspace{-.5in} \hspace{-1.5in} \vbox{
\begin{flushleft}
 $^{\textrm{\normalsize
(1)\ Department of Physics, Ben-Gurion University,
   Beer-Sheva 84105, Israel}}$
$^{\textrm{\normalsize (2)\ Department of Physics \& Electronics, Rhodes University,
 Grahamstown 6140, South Africa}}$
$^{\textrm{\normalsize (3)\ National Institute for Theoretical Physics (NITheP), Western Cape 7602,
South Africa}}$
\\ \small \hspace{0.57in}
   ramyb@bgu.ac.il,\  j.medved@ru.ac.za,\ tomshin@post.bgu.ac.il
\end{flushleft}
}}
\date{}
\maketitle

\renewcommand{\baselinestretch}{1.15}

\begin{abstract}

  The frozen star model provides a classical description of a regularized black hole and is based upon the idea that regularizing the singularity requires deviations from the Schwarzschild geometry which extend over horizon-sized scales, as well as maximally negative radial pressure as an equation of state.  The frozen star has also been shown to be ultra-stable against perturbations; a feature that can be  attributed to the equation of state and corresponds to this model  mimicking a black hole in the limit $\hbar\to 0$ or, equivalently, the limit of infinite Newton's constant.  Here, we  ``defrost'' the frozen star by allowing its radial pressure to be perturbatively less negative than maximal.  This  modification to  the equation of state is implemented by  appropriately  deforming the background metric so as to  allow the frozen star to mimic a quantum black hole at finite $\hbar$ and Newton's constant. As a consequence, the defrosted star acquires a non-trivial spectrum of oscillatory perturbations.  To show this, we first use the Cowling approximation to obtain generic equations for  the energy density and pressure perturbations of a static, spherically symmetric background with an anisotropic fluid. The particular setting of a deformed  frozen star is then considered, for which the dispersion relation is obtained to leading order in terms of  the deviation from maximal pressure. The current results compare favorably with those obtained earlier  for the collapsed polymer model, whose strongly non-classical interior is argued to provide a microscopic description of the frozen and defrosted star geometries.
\end{abstract}
\maketitle

\newpage
\renewcommand{\baselinestretch}{1.5}
\section{Introduction}

The name ``frozen star'', which was first used by Ruffini \cite{Ruffini:1971bza} to describe objects that are now commonly known as black holes (BHs), has recently been resurrected \cite{popstar,rockstar} to describe a particular model for a regularized BH geometry \cite{bookdill,BHfollies}.  It is a highly compact object, whose compactness is parametrically close to that of a BH, and yet it manages to evade singularity theorems \cite{PenHawk1,PenHawk2} and ``Buchdahl-like'' matter bounds \cite{Buchdahl,chand1,chand2,bondi,MM} by invoking a maximally negative (radial) pressure, $\;p=-\rho\;$,~\footnote{Here, we use $p$ to denote radial pressure, $q$ for the transverse pressure components and $\rho$ is the energy density.} while not possessing a trapped surface. It also avoids certain  inconsistencies by deviating from the standard general-relativistic solution  over horizon-sized scales \cite{frolov,visser}. As far as we know, the frozen star is the first completely regular and stable solution of Einstein's equations whose compactness is arbitrarily close to that of a Schwarzschild BH. Furthermore, the solution obeys the only universally agreed-upon energy condition, the null energy condition. It is free of any pathology, unlike most, if not  all, of the other proposals for regular compact objects.

Having such a solution at hand is important as it provides a model for a compact object --- differing from a general-relativistic BH --- that can be used to estimate the validity of general relativity by looking at the resulting dynamics and  gravitational-waves emissions during a binary-BH merger event. Additionally, it provides a final state in which to address the process of gravitational collapse (a paper that includes such a discussion is already in progress).

An unusual feature of the frozen star that distinguishes  it from the gravastar \cite{MMfirst} and others of the  negative-pressure class  ({\em e.g.}) \cite{barcelo,CR} is that two of its metric component are identically vanishing, $\;g_{tt}=g^{rr}=0\;$.  However, we do not mean to  imply that  astrophysical BHs are made of an anisotropic classical fluid with maximally negative radial pressure.  The anisotropy and maximal negative pressure turned out to be needed so as to connect this solution to the collapsed polymer BH \cite{strungout}, which is another  model for a regularized compact object  that is based, in large part, on the  notion of having a maximally entropic interior \cite{inny}. The ``fluid'' that supports this maximal entropy was found to consist of long, closed, fundamental, highly excited, interacting  strings. From a geometric point of view, maximal entropy translates into a mass profile that saturates the Schwarzschild limit at all interior radii, $\;2m(r)=r\;$,  and  then $\;g_{tt}=g^{rr}=0\;$ follows from Einstein's equations. The catch is that the polymer model lacks a semiclassical geometry, as maximally entropic is synonymous with strongly non-classical. Hence, the desire  for a formal classical description, and so a frozen star model  was born. The frozen star is, therefore, merely a classical proxy of a quantum  model, the polymer BH,  which has strictly non-negative pressure. The main purpose of the frozen star model is  to facilitate some calculations that would present a challenge to someone using the  quantum model.

The frozen star and polymer models  are then meant to provide complementary pictures of the same model for a regular, ultra-compact object. The former allows for precise calculations, whereas the latter provides  a microscopic description with connections to string theory. With that in mind, $g^{rr}$ and $|g_{tt}|$ should be small enough to maintain the frozen stars' character,  but  not that much changes if these components are alternatively assigned a small but finite constant, which we denote as $\;\varepsilon\ll 1\;$. What is maintained is the frozen  star's ultra-stability against small perturbations of the metric and fluid densities \cite{bookdill,popstar}. This stability is due to the equation of state $\;p+\rho=0\;$. To see this, one can inspect the stability calculation in \cite{popstar}, where the vanishing of the metric components (but not the equation-of-state relation) was relaxed for a portion of the interior. Additionally, the limit $\;p+\rho \to 0\;$ can be taken in the current analysis to verify this conclusion. This ultra-stability was also found for the polymer model in the absence of quantum effects, which  corresponds in that case  to the absence of closed-string coupling or $g_s^2$ corrections \cite{emerge,collision}. The reason being that, in the limit $\;g_s^2 \to 0\;$ or, equivalently, $\;\hbar\to 0\;$, the  polymer model  behaves just like  a hairless classical BH. It was also shown in \cite{collision} that, once stringy quantum effects are taken into account, the polymer model does exhibit  a spectrum of excitations and some quantum hair.

The main objective of this paper is to determine the dispersion relation for the oscillatory modes of a appropriately deformed version of the frozen star; the
``defrosted star''. To have such modes at all, it is clear that the background must
be suitably modified to support fluctuations. Technically speaking, this means that the original equation of state needs to be changed such that the radial pressure is no longer exactly equal  to the negative of  the energy density and the transverse pressure components ---  which are  identically zero when the star is truly frozen ---   no longer vanish. The change that we consider is parametrically small and is supposed to model some microscopic corrections to the equation of state of the constituents. As just discussed, in the polymer model, these
corrections can be attributed to  quantum perturbations that are proportional to the square of
the string coupling strength \cite{collision}.

The spectrum of oscillatory modes for the defrosted  star could be relevant to  BH merger events. If astrophysical BHs are indeed described by such objects, the internal modes of the star could be excited during a  merger event and then act as sources for gravitational-wave emissions.  After all, it is the success of the LIGO and Virgo collaboration at detecting and then analyzing  the waves from such mergers --- starting with the now world-famous event GW150914 \cite{LIGO} and the data analysis thereof \cite{LIGO1} --- that  has led to an intensive theoretical effort to understand  how  departures from general relativity would imprint on the gravitational-wave spectrum of a BH. In this regard, each candidate for a regular BH mimicker can be expected to have its own unique and experimentally verifiable or falsifiable signature (see \cite{carded} for further discussion and  a catalogue of relevant models).

We also have in mind a previous study that determined the mode spectrum for the collapsed polymer model \cite{collision} (also see \cite{spinny}). We would like to verify that the two models produce compatible results, even though the polymer analysis  was, at times,  heuristic in nature because of the absence of an interior geometry. It is already notable that, in both cases, non-trivial mode solutions require some out-of-equilibrium physics \cite{ridethewave}.

The spectral analysis for the collapsed polymer model utilized the Cowling approximation \cite{cowling}, which assumes, for a calculation of the fluid-mode spectrum, that the spacetime modes have fully decoupled. For the sake of consistency and simplicity, we will use the same approximation here. Although it is, in any event, a justifiable assumption insofar as the Cowling approximation only gets more accurate as the star gets more compact \cite{KSmodel}.

The rest of the paper is arranged as follows: We start by deriving the required perturbative  formalism for an anisotropic fluid, with our only assumptions  being that the background geometry is static and spherically symmetric.~\footnote{We also insist on $3+1$ spacetime dimensions for concreteness.} This part of the analysis culminates in a coupled pair of equations that  describe the dynamics of two types of modes: one is associated with fluctuations of the radial  velocity and the other with those of the transverse velocity.  Some brief background material on the frozen star geometry is then provided; after which the  deformations  are introduced and explained in some detail. We then return to the dynamical equations for the modes but now with the focus on the solutions for the
defrosted or deformed frozen star. The  scaling behavior of the modes and their dispersion relation are identified and discussed.
The paper ends with a brief overview followed by an appendix that explains in a thorough way how to obtain one of the key equations in the generic portion of the perturbative analysis.

\section{Perturbation equations: Generic formalism}

In the following, the perturbation equations for the energy density and pressure of a star containing anisotropic fluid are reproduced  in the Cowling approximation, meaning that the fluctuations
of the metric are neglected. We will be closely following and reviewing the analysis of  \cite{showerheads}, while  providing additional details as  needed to ensure compatibility of the generic formalism  with that of the defrosted star model.

\subsection{Variation of the stress tensor}

The background under consideration is, for the meantime, described by
a static, spherically symmetric but otherwise generic geometry,
\begin{equation}
ds^{2}\;=\;-e^{2\Phi(r)}dt^{2}+e^{2\Lambda(r)}dr^{2}+r^{2}d\Omega^{2}\;.
\label{linele}
\end{equation}
 The corresponding stress (energy--momentum) tensor can be expressed as
\begin{equation}
T_{\mu\nu}\;=\;\rho u_{\mu}u_{\nu}+pk_{\mu}k_{\nu}+q\left(g_{\mu\nu}+u_{\mu}u_{\nu}-k_{\mu}k_{\nu}\right)\;,\label{SET}
\end{equation}
where the transverse pressure $q$ is generally different from the radial pressure
$p$, so that  $\;\sigma=p-q\neq 0\;$.
Also, $u^{\mu}$ is the fluid 4-velocity  with standard normalization, $\;u^{\mu}u_{\mu}=-1\;$,
 and $k^{\mu}$ is a
radial unit vector  that is defined  by $\;k^{\mu}k_{\mu}=+1\;$ such that $\;u^{\mu}k_{\mu}=0\;$.

The oscillatory modes of the fluid will be obtained from the
perturbation equations that arise  from the variation of the conservation
equation $\;\nabla_{\nu}T_{\mu}^{~\nu}=0\;$. In the Cowling approximation, the
variation of interest is
\begin{equation}
\nabla_{\nu}\delta T_{\mu}^{~\nu}\;=\;0\;,\label{varT}
\end{equation}
 and the variation of Eq.~(\ref{SET}) under this
approximation yields
\bea
\delta T_{\mu}^{\nu}\;=\;&&\left(\delta\rho+\delta q\right)u_{\mu}u^{\nu}+\left(\rho+q\right)\left(\delta u_{\mu}u^{\nu}+u_{\mu}\delta u^{\nu}\right) \cr &+&\delta q\delta_{\mu}^{\nu}+\delta\sigma k_{\mu}k^{\nu}+\sigma\delta k_{\mu}k^{\nu}+\sigma k_{\mu}\delta k^{\nu}\;.
\label{varT2}
\eea

The velocity $u^{\mu}$ satisfies some useful identities,
\begin{equation}
\nabla_{\nu}u^{\nu}\;=\;u^{\mu}\nabla_{\nu}u_{\mu}\;=\;u^{\mu}\delta u_{\mu}\;=\;0\;,
\label{useful}
\end{equation}
and similarly for the radial unit vector  $k^{\mu}$.
By choosing, without loss of generality, to work  in the comoving  frame,
we also have
$\;u^{\mu}=u^{t}\delta_{t}^{\mu}\;$, $\;k^{\mu}=k^{r}\delta_{r}^{\mu}\;$
and $\;u^{t}u_{t}=-k^{r}k_{r}=-1\;$.

Equation~(\ref{varT}) leads to  an independent pair of perturbation equations when
it is projected, respectively,
parallel and perpendicular to $u^{\mu}$; the latter  by way of  the projection
operator $\;\mathcal{P}_{\alpha}^{\mu}=\delta_{\alpha}^{\mu}+u^{\mu}u_{\alpha}\;$.
Taking the covariant derivative of Eq.~(\ref{varT2}) and projecting onto
$u^{\mu}$, one obtains a scalar equation of the form
\begin{align}
0\;=\;u^{\mu}\nabla_{\nu}\delta T_{\mu}^{\nu}\;=\;& -\nabla_{\nu}\delta\rho u^{\nu}-\nabla_{\nu}\left[\left(\left(\rho+q\right)\delta_{\mu}^{\nu}+\sigma k^{\nu}k_{\mu}\right)\delta u^{\mu}\right]\nonumber \\
 & -\left(\rho+q\right)a_{\mu}\delta u^{\mu}-\nabla_{\nu}u^{\mu}\delta\left(\sigma k^{\nu}k_{\mu}\right)\;,\label{udeltaT}
\end{align}
where $\;a_{\mu}=u^{\nu}\nabla_{\nu}u_{\mu}\;$ is the 4-acceleration.
Whereas, in the direction perpendicular to $u^{\mu}$, one
obtains the following vector equation  (see Appendix~A for details):
\begin{align}
0\;=\;\mathcal{P}_{\alpha}^{\mu}\nabla_{\nu}\delta T_{\mu}^{\nu}\;=\; & \delta\left(\rho+q\right)a_{\alpha}+\left(\rho+q\right)u^{\nu}\left(\nabla_{\nu}\delta u_{\alpha}-\nabla_{\alpha}\delta u_{\nu}\right)\nonumber \\
 & +\nabla_{\nu}\delta q\delta_{\alpha}^{\nu}+u^{\nu}u_{\alpha}\nabla_{\nu}\delta q+\mathcal{P}_{\alpha}^{\mu}\nabla_{\nu}\delta\left(\sigma k_{\mu}k^{\nu}\right)\;.\label{PdeltaT}
\end{align}

\subsection{Density and pressure perturbations}

It proves to be useful if the perturbations of the spacelike components of the velocity vector  are expressed in terms of a displacement
vector $\xi^{i}$ such that
\begin{equation}
\frac{\partial\xi^{i}}{\partial t}\;=\;\frac{\delta u^{i}}{u^{t}}={ v}_i\;,
\end{equation}
where $\;i=\{r, \theta, \phi\}\;$ and $v_i$ is a component of the fluid's 3-velocity.

As the velocity is a function of $t$ and $r$ only,  $\;a_{\theta}=a_{\phi}=0\;$.
Then  the angular components of Eq.~(\ref{udeltaT})  reduce to
\begin{equation}
\left(\rho+q\right)\left(u^{t}\right)^{2}\partial_{t}^{2}\xi_{\theta}+\partial_{\theta}\delta q\;=\;0\;,
\end{equation}
\begin{equation}
\left(\rho+q\right)\left(u^{t}\right)^{2}\partial_{t}^{2}\xi_{\phi}+\partial_{\phi}\delta q\;=\;0\;.
\end{equation}
Differentiating  the former equation by $\phi$  and the latter  by $\theta$, one
is led to a consistency condition,
\begin{equation}
\partial_{\phi}\xi_{\theta}\;=\;\partial_{\theta}\xi_{\phi}\;,
\end{equation}
which implies that $\xi_{\theta}$ and $\xi_{\phi}$ can be expressed as  angular derivatives of a multipole expansion \cite{ThorneRMP},
\begin{align}
\xi_{\theta} & \;=\;-\sum_{\ell,m}V_{\ell m}\left(r,t\right)\partial_{\theta}Y_{\ell m}\left(\theta,\phi\right)\;,
\label{formth} \\
\xi_{\phi} & \;=\;-\sum_{\ell,m}V_{\ell m}\left(r,t\right)\partial_{\phi}Y_{\ell m}\left(\theta,\phi\right)\;.
\label{formph}
\end{align}

The definition of $\xi^{i}$ enables one to  obtain the variation in the energy density by
integrating Eq.~(\ref{udeltaT}) over time, which eventually leads to
\begin{align}
\delta\rho\;=\; & -\frac{1}{\sqrt{-g}}\partial_{\nu}\left[\sqrt{-g}\left(\left(\rho+q\right)\delta_{i}^{\nu}+\sigma k^{\nu}k_{i}\right)\xi^{i}\right]\nonumber \\
 & -\left[\left(\left(\rho+q\right)\delta_{i}^{\nu}+\sigma k^{\nu}k_{i}\right)\xi^{i}\right]\partial_{\nu}\left(\ln u^{t}\right)-\left(\rho+q\right)a_{i}\xi^{i}\;.\label{22}
\end{align}

A convenient form for the radial component of the displacement vector is the following:
\begin{equation}
\xi^{r}\;=\;\sum_{\ell,m}e^{-\Lambda}\frac{W_{\ell m}\left(r,t\right)}{r^{2}}Y_{\ell m}\;.
\label{formr}
\end{equation}
To avoid clutter, we will henceforth
be suppressing  the subscripts $\ell$ and $m$ on  $W$ and $V$,
as well as the accompanying factors of $Y_{\ell m}$ and the summation over $\ell,m$ in any equation involving $W$ and/or $V$. These indices, factors and summations can be easily restored at the end of the calculation.

We next  employ $\;u^{t}=e^{-\Phi\left(r\right)}\;$, $\;a_{\mu}=a_{r}\delta_{~\mu}^{r}=\partial_{r}\Phi\;$, the previous forms for $\xi^{i}$  and the equation for energy conservation,
\begin{equation}
\frac{dp}{dr}\;=\;-\partial_{r}\Phi\left(\rho+p\right)-\frac{2\sigma}{r}\;,
\label{energycons}
\end{equation}
to recast
Eq.~(\ref{22}) as
\begin{align}
\delta\rho\;=\; & -\left(\rho+p\right)\left[e^{-\Lambda}\frac{W^{\prime}}{r^{2}}+ \frac{\ell\left(\ell+1\right)}{r^{2}}V\right]-\frac{d\rho}{dr}e^{-\Lambda}\frac{W}{r^{2}} +\frac{2\sigma}{r^{3}}e^{-\Lambda}W\nonumber \\
 & +\sigma\frac{\ell\left(\ell+1\right)}{r^{2}}V
\;, \label{deltarho}
\end{align}
where a prime denotes a differentiation by $r$.

 The perturbation of the
radial pressure can be  obtained by first recalling the relation between the Eulerian ($\delta\rho$)
and Lagrangian ($\Delta\rho$) variations of the energy density,
\begin{equation}
\delta\rho\;=\;\Delta\rho-\xi^{r}\partial_{r}\rho\;.
\end{equation}
Given an equation of state   of the form $\;p=p\left(\rho\right)\;$,
the  Lagrangian perturbations for the radial pressure and density  can be directly related,
\begin{equation}
\Delta p\;=\;\frac{dp}{d\rho}\Delta\rho\;=\;\frac{dp}{d\rho}\left(\delta\rho+\xi^{r}\partial_{r}\rho\right)\;,
\end{equation}
and the Eulerian variation of the pressure goes similarly as
\bea
\delta p &=&\Delta p-\xi^{r}\partial_{r}p \nonumber \\
&=&\frac{dp}{d\rho}\left(\delta\rho+\xi^{r}\partial_{r}\rho\right)-\xi^{r}\partial_{r}p \nonumber \\
&=&  \frac{dp}{d\rho} \;\delta\rho\;, \label{Euly}
\eea
where the last line follows from  $\;\left(dp/d\rho\right)\partial_{r}\rho=\partial_{r}p\;$.

More explicitly,
\begin{align}
\hspace{-0.5in}\delta p\;=\; & \frac{dp}{d\rho}\left(-\left(\rho+p\right) \left[e^{-\Lambda}\frac{W^{\prime}}{r^{2}}+ \frac{\ell\left(\ell+1\right)}{r^{2}}V \right] +\frac{2\sigma}{r^{3}}e^{-\Lambda}W+\sigma\frac{\ell\left(\ell+1\right)}{r^{2}}V
\right)\nonumber \\
 & -\frac{dp}{dr}e^{-\Lambda}\frac{W}{r^{2}}\;.
 \label{deltap}
\end{align}

Although the Eulerian variation $\delta p$ is the quantity that determines the form of the oscillatory modes, the Lagrangian variation $\Delta p$ is the quantity that is directly relevant to the boundary conditions. One can see from Eq.~(\ref{formr}) and the first line of Eq.~(\ref{Euly}) that  the difference between the two variations is the same as the sole term in the second line of Eq.~(\ref{deltap}).  Hence, the Lagrangian variation $\Delta p$ can be obtained directly from Eq.~(\ref{deltap}) by simply dropping the second line.

\subsection{Dynamical equations for oscillatory modes}

We are interested in oscillating  modes of the form $\;W\left(r,t\right)=W\left(r\right)e^{i\omega t}\;$
and likewise for $V$. The dynamical equations for $W$ and $V$ follow from
Eq.~(\ref{PdeltaT}) for $\;\mu=r\;$  and $\;\mu=\theta\;$, respectively, and take on the
corresponding forms,
\begin{align}
0\;=\; & -\omega^{2}\left(\rho+p\right)e^{\Lambda-2\Phi}\frac{W}{r^{2}}+\partial_{r}{\delta p}+\left({\delta\rho}+{\delta p}\right)\partial_{r}\Phi + \frac{2}{r}{\delta \sigma}
\;\label{dynamicW}
\end{align}
and
\begin{equation}
0\;=\;\left(\rho+q\right)e^{-2\Phi}\omega^{2}V+{\delta q}\;,
\label{dynamicV}
\end{equation}
where, just like for the $W$ and $V$ modes, the   $\ell$, $m$  indices on
$\delta \rho$, the associated summations and  accompanying spherical harmonics are all implied in the above and subsequent equations. So $\delta\rho$ here is meant to represent  ${\delta\rho}_{\ell m}Y_{\ell m}$ such that   the total Eulerian density perturbation is given by
$\;\sum\limits_{\ell,m}{\delta\rho}_{\ell m}Y_{\ell m}\;$.

The next step in \cite{showerheads} and similar discussions  relies  on a formal equation-of-state relation between $\rho$ and $p$ and also between  $q$ (or $\sigma$) and $p$, as these enable  one to calculate
$\frac{\partial \rho}{\partial p}$ and then $\;\delta \rho=\frac{\partial \rho}{\partial p} \delta p\;$ (and similarly for $\delta q$).
For the frozen star model,  one has  $\;\rho+p=0\;$  but no such  relation exists  between $q$ and $p$ because the transverse pressure  is  identically vanishing.  Nevertheless, the
situation changes for the defrosted star, as  the $\rho$--$p$ relation
picks up a perturbative correction and we are now able to relate
$q$ and $p$
by varying (in an Euler--Lagrange sense)  the conservation equation~(\ref{energycons}), which leads to
$\;\frac{\partial q}{\partial p}\;$ and thus $\;\delta q=\frac{\partial q}{\partial p} \delta p\;$.

Expressing  ${\delta \rho}$ and ${\delta q}$  in terms of ${\delta p}$ and background quantities allows us to rewrite both   of the dynamical equations~(\ref{dynamicW},\ref{dynamicV}) in terms of
$\delta p$,  which itself is determined by Eq.~(\ref{deltap}). The final result is then  a pair of  coupled equations in terms of a pair of unknown functions, $W(r)$ and $V(r)$, for any given values of $\ell$ and $m$. Thanks to the appearances of $\partial_{r}{\delta p}$ in Eq.~(\ref{dynamicW}) and $W'$ in Eq.~(\ref{deltap}), one can, after diagonalizing,  expect to obtain a second-order differential
equation for $W(r)$ and first-order equation for $V(r)$.

 The dynamical  equations are to be supplemented by a pair of boundary conditions
involving the Lagrangian variation $\Delta p$  of the radial pressure. In particular, the Lagrangian variation should vanish both at the star's outer surface $\;r=R\;$ and
at the star's center $\;r=0\;$. Importantly, this variation should
also remain finite as it approaches  the center of the star;
that is, for $\;r \ll R\;$.

\section{Frozen star: A refresher}

To formulate the frozen star model, one starts with a metric and stress tensor just like
those
in Eqs.~(\ref{linele}) and~(\ref{SET}).
The first key ingredient is that the interior is endowed with  a maximally negative radial pressure, $\;p=-\rho\;$, throughout.
Consistency between the Einstein field equations and the  energy-conservation equation~(\ref{energycons}) is what
dictates the form of the transverse pressure $q$.

Next, let us define a mass function $m(r)$ in the standard way,
\be
\label{mofr}
m(r)\;=\;4\pi\int\limits_0^r dx\, x^2 \rho(x)\;\;\; {\rm for} \;\;\; r\leq R\;,
\ee
for which Einstein's equations indicate that
\be
e^{-2\Lambda} \;=\; 1- \frac{2 m(r)}{r}\;.
\ee

It is the choice of $m(r)$ that specifies the exact nature of the model within this negative-pressure  class.
 The most popular option is to choose
 $m$ such  that $\rho$ is constant, with the result that $\;q=p=-\rho\;$; what  is  known as the gravastar model \cite{MMfirst}.
In  its originally prescribed form, the frozen star was specified by the profile
$\;m(r)=r/2\;$, as this choice ensured that each spherical shell within the star saturated the Schwarzschild bound. This
would be the second key ingredient and, just like the first, it was chosen to make
the frozen star  a proxy to the (strongly non-classical) collapsed polymer model. However, as this
same mass profile also leads to a pair of  apparently singular  metric components,
$\;e^{-2\Lambda}=e^{2\Phi}=0\;$, we have more recently opted to follow a different path:
Namely, we now set $\;m(r)=\frac{r}{2}(1-\varepsilon)\;$ for a small dimensionless and  constant $\varepsilon$, and so  $\;e^{-2\Lambda}=e^{2\Phi}=\varepsilon\ll 1\;$. It is  implied that $\varepsilon$ is the smallest dimensionless scale in the model.

In this relaxed version of the frozen star, the energy density and pressure components go as
\bea
\label{polyrho}
8\pi G  \rho &=& \frac{1-(rf)'}{r^2} \;=\;  \frac{1-\varepsilon}{r^2}\;, \\
\label{polypr}
8\pi G  p &=& -\frac{1-(rf)'}{r^2} \;=\; -\frac{1-\varepsilon}{r^2}\;, \\
\label{polypt}
8\pi G  q  &=& \frac{(rf)''}{2r}\;=\; 0\;.
\eea
Not much has changed from the $\;\varepsilon=0\;$ limit. In fact,  a stability
study in \cite{popstar} reveals that the ultra-stability of the model against $r,t$-dependent perturbations persists irrespective of the inclusion of  non-zero metric components. The critical   feature for
ensuring stability is
rather the condition $\;p =-\rho\;$. Whether $\varepsilon$ is zero
or small but finite is besides the point.

Some final notes about the frozen star: It is necessary to modify the metric over a thin translational
layer near (and including) the outer boundary so that the star's metric and its first two derivatives
match smoothly to the Schwarzschild metric in the exterior \cite{popstar}. Additionally,
the metric near the center of the star needs to be regularized so as to ensure that the energy density remains finite at $\;r=0\;$ \cite{rockstar}. This regularization process  is mandatory for the frozen star because, unlike  a Schwarzschild BH, any would-be singularity is not protected by cosmic censorship.

The width of the transitional surface layer is  taken to be very small, of order of the string or Planck length scale, so that its inclusion in our analysis would lead to highly suppressed corrections. We can then safely ignore the outer layer except that its presence  is relevant to the outer-surface boundary condition.
In our case, the value of $\Delta p$ should be set equal to some non-zero value at $\;r=R\;$, thus allowing it  to  decay to zero only after passing through the
translational
layer where it would match on to the external geometry. The mathematical details of this process are inconsequential to the current analysis but would follow
along the lines of \cite{popstar}.

Similarly, the regularized core has a similarly sized width and could also  be ignored
except  for the following observation (again pertaining to boundary conditions): A particularly relevant feature of the core regularization process in \cite{rockstar} is that $\rho$, $-p$ and $-q$ all tend  to the same constant value at $\;r=0\;$, yet the rate in which they approach that value is somewhat arbitrary. We then  have the freedom to send  the combinations $\;\sigma= p-q\;$ and $\;\rho+p\;$ of the deformed metric (see the next section) to zero at an arbitrarily fast rate. This is pertinent to the boundary condition at the center because, as one can see from the first line of Eq.~(\ref{deltap}), every term  in the Lagrangian variation $\Delta p$ is preceded by just such a combination.
Meaning  that the $\;r=0\;$  regularity condition can always be easily satisfied for the defrosted star model.~\footnote{In \cite{rockstar}, the regularization method was applied to the undeformed version of the frozen star model,
but it is a straightforward exercise to
include deformations and obtain similar results.}

\subsection{Briefly on stability}

Before discussing the deformation of the frozen star, let us consider how the dynamical equations~(\ref{dynamicW}) and~(\ref{dynamicV}) can be used
to learn about the star's stability in the undeformed case. Starting with  Eq.~(\ref{dynamicW}), one  immediately sees that the $W$ term vanishes
because of the factor of $\;\rho+p=0\;$ and the $a_r$ term because the equation of state dictates that $\;{\delta\rho}
+{\delta p}\;$=0 or, more simply, because $a_r$ itself is vanishing for constant $\varepsilon$.

As we are assuming an object of fixed mass,
it follows that $\;{\delta \rho}=0\;$ and then likewise for the equal and
opposite variation
$\;{\delta p}\;$.
To see that  $\;{\delta q}\;$  also vanishes, one  can
use  $\;\rho+p=0\;$ to rewrite the
conservation equation~(\ref{energycons}) as
\be
0\;=\; r^2 \partial_r p +2r \sigma\; =\;  \partial_r (r^2 p)   -2  r  q\;,
\ee
which confirms that both $q$ and   $\;{\delta q}\;$  vanish because $\;r^2 p\;$ must be constant by virtue of Eq.~(\ref{polypr}) and $\delta p=0\;$.

Since $\;\delta q =0\;$, Eq.~(\ref{dynamicV}) reduces to
\be
0\;=\;\left(\rho+q\right)e^{-2\Phi}\omega^{2}V\;,
\ee
meaning that either $\omega$  or $V$ is vanishing.  No matter which, the angular components of the velocity perturbations are vanishing, as these are given by the time derivative of
$V$.

This leaves the radial component of the velocity perturbations, which are determined by the time derivative of $W$. However, the dynamical equations tell us nothing about $W$ for the undeformed star because it carries a vanishing factor of $\rho+p$. The way around this is to actually deform the star (as we do next) and then consider the limit as the deformation parameter goes to zero. What one finds is that $\omega$ does indeed go to zero in this limit, meaning that the time derivative of $W$ and thus the radial component of the velocity perturbations must also tend to zero. Note, though, that it can never be established that $W$ is itself vanishing. Fortunately, there is no reason that it has to. Only the time derivative of $W$ has physical meaning.

\section{Defrosting a frozen star}

The frozen star geometry is unusually stable against perturbations, even
when $\;\varepsilon=|g_{tt}|=g^{rr}\;$ differs from zero. For the star
to sustain dynamical modes, we have found that at least two modifications are required. The first requirement is that the equation of state deviates from $\;p=-\rho\;$, which  necessitates that $\;|g_{tt}|\ne g^{rr}\;$. The second is that
at least one of $g_{tt}$ and  $g^{rr}$ have some radial dependence.
To implement these requirements, we introduce a perturbative, dimensionless and positive parameter $\;\gamma\ll 1\;$ which controls the difference between the frozen and  defrosted stars.

\subsection{Deformed background geometry}

Let us first re-express the relevant metric components as
\bea
-g_{tt}&=& \varepsilon+ \gamma\left(\frac{r}{R}\right)^a\;, \\
g^{rr}&=& \varepsilon +\gamma\left(\frac{r}{R}\right)^b\;,
\eea
where  the relation between the star's outer surface $R$ and mass  $M$  will be determined shortly.
It will be shown later that the  constants $a$ and $b$ are fixed, for consistency, such that $\;a=2\;$ and $\;b=0\;$.

We assume that $\;\varepsilon\ll \gamma\;$ and so neglect $\varepsilon$ in what follows, leaving us with
\bea
-g_{tt} &=&  \gamma\left(\frac{r}{R}\right)^a\;, \\
g^{rr}&=& \gamma\left(\frac{r}{R}\right)^b\;.
\eea
Furthermore, only  the  leading-order terms in $\gamma$ will ever be considered.
We assume that regularization procedures similar to those used in \cite{popstar}
and \cite{rockstar} have been implemented but, as already discussed, these
are expected to  have no tangible affect on our analysis except to trivialize the enforcement
of the central boundary condition and modify the outer boundary condition.

Let us now consider two components  of Einstein's equations,
\be
 \;r^2 \rho  \;=\; \left[r\left(1-g^{rr}\right)\right]'\;,
\ee
\be
\;r^2 p\;=\;  g^{rr}-1 + rg^{rr}\left[\ln{|g_{tt}|}\right]'\;,
\ee
where the convention $\;8\pi G=1\;$ has now been adopted.
Then, for the deformed frozen star,
\be
 r^2 \rho \;=\;1-(1+b)~\gamma\left(\frac{r}{R}\right)^b\;,
\label{r2r}
\ee
\be
 r^2  p \;=\;-1+(1+a)~\gamma\left(\frac{r}{R}\right)^b\;,
\label{r2p}
\ee
which can be combined into
\be
p\;=\;\rho \left(-1+(a-b)~\gamma\left(\frac{r}{R}\right)^b\right)\;
\label{pvr}
\ee
and
\be
r^2  (\rho+p)\;=\; (a-b) ~\gamma\left(\frac{r}{R}\right)^b\;,
\label{r2rplusp}
\ee
thus making the deviation from $\;\rho+p=0\;$ quite clear.
Satisfying the null energy condition requires that $\;a \geq b\;$, which we will assume.

Via the energy conservation equation~(\ref{energycons}), it can be shown
that $q$ is no longer vanishing,
\be
 r^2 q \;=\; \frac{1}{4}\left(a^2+a b+2 b\right)~ \gamma\left(\frac{r}{R}\right)^b  \;,
 \label{r2q}
\ee
and consequently that
\be
  q \;=\; \frac{1}{4}\left(\frac{a^2+a b+2 b}{a-b}\right) (\rho+p)\;.  \;
 \label{r2qrplusp}
 \ee

We can use the above equation to determine $\frac{\partial p}{\partial q}$, which will be needed later on for the calculation of the  perturbation $\delta q$,
\bea
\frac{\partial q}{\partial p}&=&\frac{1}{4}\left(\frac{a^2+a b+2 b}{a-b}\right) \left(\frac{\partial \rho}{\partial p}+1\right) \cr &=& -\frac{1}{4}\left(a^2+a b+2 b\right) ~\gamma\left(\frac{r}{R}\right)^b  \;,
 \label{dpdq}
\eea
where the variation $\frac{\partial \rho}{\partial p}$ can be obtained by inverting Eq.~(\ref{pvr}).

The following relation will also be required:
\be
\partial_r \Phi\; =\;\frac{a}{r}\;.
\label{drphi}
\ee

The radius of a defrosted  star of mass $M$ is larger by a factor of order  $\frac{\gamma} R$ from its Schwarzschild size. The way to see this is to match $\;g^{rr}=\gamma\left(\frac{r}{R}\right)^b \;$ to its Schwarzschild form $\;g^{rr}=1-\frac{2GM}{r}\;$ at $\;r=R\;$, which yields $\;R= \frac{2 G M}{1-\gamma}\simeq {2 G M}(1+\gamma)\;$.
One can use this to indeed  verify that the total mass of the star is the same
mass $M$ as its undeformed counterpart. With $E$ denoting the deformed
star's mass,
\bea
E &=& \int\limits^{R}_{0} dr\; 4\pi  r^2 \; \rho(r)\; = \;\frac{1}{2G}\int\limits^{R}_{0} dr\left(1-(b+1)\gamma\left(\frac{r}{R}\right)^b\right)  \nonumber \\
&=& \frac{1-\gamma}{2G}R \;=\; M\;,
\eea
where Eq.~(\ref{r2r}) (with a restored factor of $8\pi G$ on its left side)  has been used in the top line.

\subsection{Perturbations}

We start this discussion by recalling the pressure perturbation in Eq.~(\ref{deltap}),
\bea
r^2 \delta p\;=\;  (\rho+p)  e^{-\Lambda}{W^{\prime}}-\left[\partial_r(r^2 p)-2 r q \right] e^{-\Lambda}\frac{W}{r^{2}}
-~p~\ell\left(\ell+1\right){V}\;.
\label{deltapFS1}
\eea
To obtain this form from the previous one, we have used that the factors in front of $  e^{-\Lambda}W'$ and  $ e^{-\Lambda}W$ are
linear in $\gamma$ (the latter by using the conservation equation) and that, as shown later,~\footnote{See the scaling relations that immediately follow
Eq.~(\ref{wtilde}).} the leading-order contribution to
$V$ is also linear in $\gamma$. These observations allowed us, in particular,  to set $\frac{\partial p}{\partial \rho}=-1\;$, recalling that our calculations are performed to leading order in $\gamma$.

Substituting the background quantities for the deformed geometry from the previous section, we find that, to leading order in $\gamma$,
\bea
r^2 \delta p\;=\; (a-b)\gamma \left( \frac{r}{R}\right)^{b}~\left[e^{-\Lambda}\frac{W^{\prime}}{r^2} +\frac{a}{2}e^{-\Lambda} \frac{W}{r^3}\right]
+~\ell\left(\ell+1\right)\frac{V}{r^{2}}\;.
\label{deltapFS2}
\eea

The two other perturbations $\delta \rho$ and $\delta q$ can be related to $\delta p$ by using Eqs.~(\ref{pvr}),~(\ref{r2rplusp}) and~(\ref{dpdq}). From Eq.~(\ref{pvr}),
\be
\delta \rho\; =\; \frac{\partial \rho}{\partial p} \delta p \;\simeq\; - \delta p\;,
\label{deltar}
\ee
and then from Eq.~(\ref{r2rplusp}),
\be
\delta \rho+  \delta p \;\simeq\; -(a-b)\gamma \left( \frac{r}{R}\right)^{b} \delta p\;,
\label{deltarp}
\ee
and finally, from Eq.~(\ref{dpdq}),
\be
\delta q \;=\; \frac{\partial q}{\partial p}\delta p \;\simeq\; -\frac{1}{4}\left(a^2+a b+2 b\right) \gamma \left( \frac{r}{R}\right)^{b} \delta p \;.
\label{deltaq}
\ee
As expected, the perturbation in the transverse pressure is subleading to
the radial perturbation, $\;\delta q\sim \gamma~ \delta p\;$, same as for the deformed background components,
$\;q\sim \gamma ~p\;$.

\section{Frozen star: Dynamical equations and dispersion relations}

The initial step  here is to rewrite the dynamical equations~(\ref{dynamicW},\ref{dynamicV}) for $W$ and $V$ in terms of the deformed frozen star geometry.
Recall the suppression of the angular indices, accompanying spherical harmonics
and summations.

First, multiplying  Eq.~(\ref{dynamicW}) by $r^2$ and then rewriting,
we find that
\bea
&-&\frac{\omega^2}{\gamma^2 \left( \frac{r}{R}\right)^{a+b} }
\left((a-b) ~\gamma\left(\frac{r}{R}\right)^b\right) e^{-\Lambda}\frac{W}{r^{2}} \cr  &+&\partial_{r}(r^2 \delta p) +\frac{1}{2} a r \left({\delta\rho}+{\delta p}\right)  -2 r\delta q \;=\; 0\;.
\;\label{dynamicWFS}
\eea
Meanwhile, for Eq.~(\ref{dynamicV}), the resulting expression is
\begin{equation}
 \frac{\omega^2}{\gamma^2 \left( \frac{r}{R}\right)^{a} }\; \gamma\;\frac{V}{r^2}+{\delta q}\;=\;0\;.
\label{dynamicVFS}
\end{equation}

We may now substitute the expressions for $\delta q$ from Eq.~(\ref{deltaq}) and for
${\delta\rho}+{\delta p}$ from Eq.~(\ref{deltarp}) into
the two previous equations, giving
\bea
\label{dynamicWFS1}
&-&\frac{\omega^2}{\gamma^2 \left( \frac{r}{R}\right)^{a+b} }
\left((a-b) ~\gamma\left(\frac{r}{R}\right)^b\right)
e^{-\Lambda}\frac{W}{r^{2}} \\  &+&\partial_{r}(r^2 \delta p) - \frac{1}{2}a(a-b)\gamma \left( \frac{r}{R}\right)^{b} r \delta p  +\frac{1}{2}\left(a^2+a b+2 b\right) \gamma \left( \frac{r}{R}\right)^{b} r \delta p\; =\; 0 \nonumber
\;
\eea
and
\begin{equation}
\frac{\omega^2}{\gamma^2 \left( \frac{r}{R}\right)^{a} }\; \gamma\; \frac{V}{r^2} - \frac{1}{4}\left(a^2+a b+2 b\right) \gamma \left( \frac{r}{R}\right)^{b} \delta p\;=\;0\;.
\label{dynamicVFS1}
\end{equation}

An inspection of  Eq.~(\ref{deltapFS2}) reveals  the scaling $\;r^2 \delta p \sim  V/r^2 \;$, whereas
Eq.~(\ref{dynamicVFS1}) rather implies $\; r^{a+b} \delta p \sim V/r^2\;$.
The conclusion is that
\be
a+b\;=\;2\;.
\label{aplusb}
\ee

Similarly, by comparing the radial dependence
of the term $\partial_{r}(r^2 \delta p)$ in Eq.~(\ref{dynamicWFS1}) with any of the other terms involving $\delta p$, one observes that
$\;r\sim r^{b+1}\;$, so that
\bea
a&=&2\;, \\
b&=&0\;.
\label{ab}
\eea

These values for $a$ and $b$  simplify the dynamical equations and the expression for  $\delta p$ by a considerable amount. Respectively,
\bea
  &-&2\frac{ \omega^{2} R^2}{\gamma^2} {\gamma}^{3/2}\frac{W}{r^4}   + \partial_{r} (r^2 \delta p) \; =\; 0\;,
\;\label{dynamicWFS2}
\eea
\begin{equation}
 \frac{\omega^{2}R^2}{\gamma^2} \frac{V}{r^2} -  r^2 \delta p\;=\;0\;,
\label{dynamicVFS2}
\end{equation}
\bea
r^2 \delta p\;=\; 2\gamma^{3/2} \left[\frac{ W^{\prime}}{r^2} + \frac{ W}{r^3}\right]
+~\ell\left(\ell+1\right)\frac{V}{r^{2}}\;.
\label{deltapFS3}
\eea

Defining a dimensionless frequency $\widetilde{\omega}$,
\be
\omega^2 \;=\; \gamma^2 \tfrac{1}{R^2} \widetilde{\omega}^2
\label{wtilde}
\ee
and  rescaling the perturbations as $\;\widetilde{W}=\sqrt{\gamma} W\;$, $\gamma\widetilde{V}= V\;$ and $\;{\gamma} \widetilde{\delta p}=r^2{\delta p}\;$,
we then have
\bea
  \partial_{r}  \widetilde{\delta p} \; =\; 2 \widetilde{\omega}^{2}\frac{\widetilde{W}}{r^{4}} \;,
\;\label{dynamicWFSFF}
\eea
\begin{equation}
 \widetilde{\delta p}\;=\; \widetilde{\omega}^{2}  \frac{\widetilde{V}}{r^2}\;,
\label{dynamicVFSFF}
\end{equation}
\bea
\widetilde{\delta p}\;=\; 2 \frac{\partial_r \left(r\widetilde{W}\right)}{r^3}
+~\ell\left(\ell+1\right)\frac{\widetilde{V}}{r^{2}}\;.
\label{deltapFSFF}
\eea

We conclude that the the oscillation frequencies are non-relativistic, governed by a scale of
$\gamma/R$ which is parametrically smaller than the relativistic frequency
scale; that is,
$\;\omega\sim \frac{\gamma}{R} \ll \frac{1}{R}\;$.

We can use Eqs.~(\ref{dynamicWFSFF}-\ref{deltapFSFF}) to obtain a single ``wave equation''   for $\widetilde{\delta p}$ (or, equivalently, for $\widetilde{W}$ or $\widetilde{V}$),
\be
-\widetilde{\omega}^2\widetilde{\delta p}  +\frac{1}{r^3} \partial_r\left(r^5 \partial_r \widetilde{\delta p}\right)+\ell (\ell+1) \widetilde{\delta p} \;=\;0\;.
\label{dynamicdpF}
\ee

To find the spectrum resulting from Eq.~(\ref{dynamicdpF}), we need to impose
a new  boundary condition to replace the one at the center of the star. Recall
that the center-of-the-star condition is trivially satisfied in our case
because of the regularization process in the core.
The new  condition is set by the scaling of the radial dependence
of the perturbations modes $V_{\ell m}$ and $W_{\ell m}$. This will be applied for  $\;r\ll R\;$, but far enough from the center so that $r$ falls within the bulk of the frozen star, outside
of the regularized core.

Let us first  consider  the expansion of $V_{\ell m}$ in Eqs.(\ref{formth}) and (\ref{formph}).
As in the multipole expansion of any such quantity,
\be
V_{\ell m} \;\sim\; V_{\ell}\; r^{\ell}\;,
\ee
where the  relevant expansion range is  $\;\ell\geq 2\;$.

Now combining Eqs.~(\ref{dynamicWFSFF}) and (\ref{dynamicVFSFF}),
we obtain
\be
\partial_r\left(\frac{\widetilde{V}_{\ell m}}{r^2}\right)\;=\; \frac{2}{r^4} \widetilde{W}_{\ell m}\;,
\ee
from which it follows that
\be
\widetilde{W}_{\ell m} \;\sim\; \widetilde{W}_{\ell}\; r^{\ell+1}
\ee
and then
\be
\widetilde{W}_{\ell m} \;=\; \frac{\ell-2}{2} \widetilde{V}_{\ell m}\;.
\ee
Additionally, it follows that $\; \widetilde{\delta p} \sim r^{\ell-2}\;$.

Using these scaling relations, we  observe that
\be
\frac{1}{r^3} \partial_r\left(r^5 \partial_r\widetilde{\delta p}\right)
\;=\; (\ell+2)(\ell-2) \widetilde{\delta p}\;,
\label{dpeqF}
\ee
and then conclude from this equation and  Eq.~(\ref{dynamicdpF}) that
\be
\widetilde{\omega}^2\;=\;2\ell^2 + \ell -4\;.
\label{w2l}
\ee

Assuming  a wavelength of order $R$, we can also deduce a speed of propagation (squared) directly from the spectrum,   $\; v^2 \sim \gamma^2 (2\ell^2 + \ell -4) \;$.
However,  from an {\em internal} (I) or proper-time perspective, one factor of the redshift $\gamma$ should be dispensed with, leaving  $\; v_I^2 \sim \gamma (2\ell^2 + \ell -4)\;$.
The dispersion relation~(\ref{w2l}) is in  agreement with that of \cite{collision}, which considered the quasinormal-mode problem for the closely related polymer model. There,  the square of the real part of the frequencies scaled with the  coupling strength of  closed strings $g_s^2$;  the
smallest non-classical, dimensionless  parameter in the framework and
the smallest dimensionless one besides $\;\epsilon=l_s/R\;$. Now recall that $\gamma$ played precisely the same role for the deformed frozen star, as $\varepsilon$ --- the redshift of the undeformed star ---  is the only smaller dimensionless parameter. Another way of seeing that $\gamma$ and $g_s^2$ play the same role in their respective models is that they both provide a direct measure of $\frac{\Delta R}{R}$. From
the polymer point of view, this relative scaling is already clear
in \cite{collision},  but see also \cite{QLove,CLove} for further discussion.

The solutions for the perturbations can be obtain  by  using Eq.~(\ref{dpeqF}) to solve  for $\widetilde{\delta p}$,
\be
\widetilde{\delta p}_\ell \;=\; C_1~ r^{l-2} + C_2~ r^{-(l+2)}\;.
\ee
The boundary condition at $\;r \ll R\;$  requires that  $\;C_2=0\;$, so that
\be
\widetilde{\delta p}_\ell \;=\; C_1~ r^{l-2}\;.
\ee
Then, from Eq.~(\ref{dynamicVFSFF}), it follows that
\be
\widetilde{V}_\ell\;=\; C_1~ \frac{1}{\widetilde{\omega}^2} ~ r^l \;,
\ee
whereas Eq.~(\ref{dynamicWFSFF}) leads to
\be
\widetilde{W}_\ell=C_1~ \frac{l-2}{2}~ \frac{1}{\widetilde{\omega}^2}~ r^{l+1}.
\ee
Note that  $\widetilde{W}_\ell$ vanishes at leading order for $\;\ell =2\;$.

An interesting feature of the spectrum in \cite{collision} was that the imaginary part of the mode frequencies ({\em i.e.}, the inverse of the damping time)  scaled as $v^2$, and so $\;{\Im}(\omega)
\ll {\Re}(\omega)\;$. This behavior is not obvious at this point in the current analysis.
But this is  because  we have yet to provide a mechanism for any of the modes to couple to and then escape into the external spacetime, which is what is required for the modes to dissipate energy and thus experience damping. However, it is easy to see why such a scaling is an inevitable feature of any ultra-compact object  that is using non-relativistic fluid modes to source gravitational waves. As explained in \cite{ridethewave}, from an external observer perspective, these modes  have exceptionally long wavelengths $\;\lambda \sim R/v\;$ when compared to the size $R$ of the object, reducing the transmission cross-section through a surface of area $R^2$ by a factor of $\;\lambda^2/R^2\sim v^2 \;$. This cross-section determines, in turn, the power loss $\;\frac{dE}{dt}\sim\lambda^2/R^2\sim v^2 \;$, which then sets the damping time scale as $\;1/\tau \sim v^2\;$, and so $\;{\rm Im}(\omega) \sim v^2\;$.

\section{Overview}

We have calculated the oscillatory mode spectrum for a deformed version of the  frozen star model: the defrosted  star. The deformations were a mathematical necessity; otherwise, the ultra-stability of the frozen star geometry would have doomed the perturbative modes from the get go. The deformations do, however, make sense when one considers that gravitational waves are generally emitted as the result of some out-of-equilibrium event like the merger of a binary system. One expects the disturbed object to eventually settle back into its original equilibrium state. It would be interesting to understand how this process
works in the case of a frozen star.

We have found a spectrum for the star that is in agreement with that of the polymer model.
In both cases, the square of the real part of the frequency  scales with a small dimensionless parameter; the deformation parameter $\gamma$ in the case of the defrosted version of the frozen star and the closed-string coupling strength $g_s^2$ for the collapsed polymer. These are seemingly very different but have two important similarities: (1) They  both represent  the second-smallest dimensionless  parameter in their respective frameworks  and the smallest one that is also  non-classical (effectively so, in the case of the star)  and (2) they both scale as $\frac{\Delta R}{R}$. We expect that this is really the same parameter but  different observers will assign it a different meaning depending on their vantage point. For instance, anyone who believes that BHs are purely classical, geometric objects would have no room for string coupling in her ``story''. This reasoning falls very much in line with what Hawking called the ``Principle of Ignorance'' \cite{Haw}.

Missing in our analysis is a direct calculation of the damping time. This could  be done by changing the boundary conditions at the surface of the star from a vanishing wave to a standing wave, which is then matched to an outgoing wave in the exterior. This calculation is technically quite involved and requires
the incorporation of  the metric perturbations, meaning that  the Cowling approximation can no longer be applied.  We do hope to perform such a calculation in the future. As explained in the main text,  it can be expected on physical grounds  that the damping time scales as $1/\gamma^2$, a result that was indeed substantiated in \cite{collision}. This is well worth confirming for the defrosted  star, as the implication is a particularly long lifetime for the emitted modes and, with it, an excellent chance of  detection. Moreover, if $\gamma$ is indeed related to the coupling strength of closed strings, the lifetime of the fluid modes provides an unexpected  window into fundamental string theory.

\newpage

\section*{Acknowledgments}
We thank Daniela  Doneva and Stoytcho  Yazadjiev for clarifying comments on their paper.
The research is supported by the German Research Foundation through a German-Israeli Project Cooperation (DIP) grant ``Holography and the Swampland'' and by VATAT (Israel planning and budgeting committee) grant for supporting theoretical high energy physics.
The research of AJMM received support from an NRF Evaluation and Rating
Grant 119411 and a Rhodes  Discretionary Grant SD07/2022.  AJMM thanks Ben Gurion University for their hospitality during his visit.

\newpage

\appendix

\section{Perpendicular projection of the perturbation \label{appendix A}}

Here, it is shown in some detail how to obtain  Eq.~(\ref{PdeltaT}), the perpendicular projection of the covariant derivative of the perturbation of the stress tensor (\ref{varT}).

Using  the properties of the 4-velocity (see Eq.~(\ref{useful})) and that  the Cowling approximation is in effect, we have
\begin{align}
\mathcal{P}_{\alpha}^{\mu}\nabla_{\nu}\delta T_{\mu}^{\nu}\;=\; & \nabla_{\nu}\left(\delta\rho+\delta q\right)\underset{=0}{\underbrace{\left(\delta_{\alpha}^{\mu}+u^{\mu}u_{\alpha}\right)u_{\mu}}}u^{\nu}\nonumber \\
 & +\left(\delta\rho+\delta q\right)\left(\left(\delta_{\alpha}^{\mu}+u^{\mu}u_{\alpha}\right)u^{\nu}\nabla_{\nu}u_{\mu}+\underset{=0}{\underbrace{\left(\delta_{\alpha}^{\mu}+u^{\mu}u_{\alpha}\right)u_{\mu}}}\overset{=0}{\overbrace{\nabla_{\nu}u^{\nu}}}\right)\nonumber \\
 & +\left(\delta_{\alpha}^{\mu}+u^{\mu}u_{\alpha}\right)\left[\nabla_{\nu}\left(\rho+q\right)\left(\delta u_{\mu}u^{\nu}+u_{\mu}\delta u^{\nu}\right)+\left(\rho+q\right)\nabla_{\nu}\left(\delta u_{\mu}u^{\nu}+u_{\mu}\delta u^{\nu}\right)\right]\nonumber \\
 & +\left(\delta_{\alpha}^{\mu}+u^{\mu}u_{\alpha}\right)\nabla_{\nu}\delta q\delta_{\mu}^{\nu}+\mathcal{P}_{\alpha}^{\mu}\nabla_{\nu}\delta\left(\sigma k_{\mu}k^{\nu}\right)\nonumber \\
 = & \left(\delta\rho+\delta q\right)\left(a_{\alpha}+u_{\alpha}u^{\nu}\underset{=0}{\underbrace{u^{\mu}\nabla_{\nu}u_{\mu}}}\right)\nonumber \\
 & +\nabla_{\nu}\left(\rho+q\right)\left(\left(\delta_{\alpha}^{\mu}+u^{\mu}u_{\alpha}\right)\delta u_{\mu}u^{\nu}+\delta u^{\nu}\underset{=0}{\underbrace{\left(\delta_{\alpha}^{\mu}+u^{\mu}u_{\alpha}\right)u_{\mu}}}\right)\nonumber \\
 & +\left(\rho+q\right)\left(\left(\delta_{\alpha}^{\mu}+u^{\mu}u_{\alpha}\right)\nabla_{\nu}\delta u_{\mu}u^{\nu}+\left(\delta_{\alpha}^{\mu}+u^{\mu}u_{\alpha}\right)\nabla_{\nu}u_{\mu}\delta u^{\nu}\right)\nonumber \\
 & +\left(\rho+q\right)\left(\delta_{\alpha}^{\mu}+u^{\mu}u_{\alpha}\right)\left(\delta u_{\mu}\overset{=0}{\overbrace{\nabla_{\nu}u^{\nu}}}+u_{\mu}\overset{=0}{\overbrace{\nabla_{\nu}\delta u^{\nu}}}\right)\nonumber \\
 & +\nabla_{\nu}\delta q\delta_{\alpha}^{\nu}+u^{\nu}u_{\alpha}\nabla_{\nu}\delta q+\mathcal{P}_{\alpha}^{\mu}\nabla_{\nu}\delta\left(\sigma k_{\mu}k^{\nu}\right)\nonumber
\end{align}
\begin{align}
\;=\; & \delta\left(\rho+q\right)a_{\alpha}+\nabla_{\nu}\left(\rho+q\right)\left(\delta u_{\alpha}u^{\nu}+u_{\alpha}\underset{=0}{\underbrace{u^{\mu}\delta u_{\mu}}}u^{\nu}\right)\nonumber \\
 & +\left(\rho+q\right)\left(u^{\nu}\nabla_{\nu}\delta u_{\alpha}+u_{\alpha}u^{\mu}u^{\nu}\nabla_{\nu}\delta u_{\mu}+u_{\alpha}u^{\mu}\nabla_{\nu}u_{\mu}\delta u^{\nu}+\nabla_{\nu}u_{\alpha}\delta u^{\nu}\right)\nonumber \\
 & +\nabla_{\nu}\delta q\delta_{\alpha}^{\nu}+u^{\nu}u_{\alpha}\nabla_{\nu}\delta q+\mathcal{P}_{\alpha}^{\mu}\nabla_{\nu}\delta\left(\sigma k_{\mu}k^{\nu}\right)\;.
\end{align}

Next, we implement our choice to work in the comoving frame, so that  $\;u^{t}u_{t}=-1\;$,
and recall the time independence of any background quantity. Then,
\begin{align}
0\;=\;\mathcal{P}_{\alpha}^{\mu}\nabla_{\nu}\delta T_{\mu}^{\nu} & =\delta\left(\rho+q\right)a_{\alpha}+\underset{=0}{\underbrace{\partial_{t}\left(\rho+q\right)}}\delta u_{\alpha}u^{t}\nonumber \\
 & +\left(\rho+q\right)\left(u^{\nu}\nabla_{\nu}\delta u_{\alpha}-u^{\mu}\overset{=\delta_{\alpha}^{\nu}}{\overbrace{\delta_{\alpha}^{t}\delta_{t}^{\nu}}}\nabla_{\nu}\delta u_{\mu}\underset{=0}{\underbrace{-\overset{=\delta_{\alpha}^{\mu}}{\overbrace{\delta_{t}^{\mu}\delta_{\alpha}^{t}}}\nabla_{\nu}u_{\mu}\delta u^{\nu}+\nabla_{\nu}u_{\alpha}\delta u^{\nu}}}\right)\nonumber \\
 & +\nabla_{\nu}\delta q\delta_{\alpha}^{\nu}+u^{\nu}u_{\alpha}\nabla_{\nu}\delta q+\mathcal{P}_{\alpha}^{\mu}\nabla_{\nu}\delta\left(\sigma k_{\mu}k^{\nu}\right)\nonumber \\
 & \;=\;\delta\left(\rho+q\right)a_{\alpha}+\left(\rho+q\right)u^{\nu}\left(\nabla_{\nu}\delta u_{\alpha}-\nabla_{\alpha}\delta u_{\nu}\right)\nonumber \\
 & +\nabla_{\nu}\delta q\delta_{\alpha}^{\nu}+u^{\nu}u_{\alpha}\nabla_{\nu}\delta q+\mathcal{P}_{\alpha}^{\mu}\nabla_{\nu}\delta\left(\sigma k_{\mu}k^{\nu}\right)\;.\label{PdeltaT-1}
\end{align}



\end{document}